# Image to Image Translation based on Convolutional Neural Network Approach for Speech Declipping

Hamidreza Baradaran Kashani
Electrical Engineering Faculty
Amirkabir University of Technology
Tehran, Iran
hr_baradaran@aut.ac.ir

Ata Jodeiri
School of Electrical & Computer Engineering
University of Tehran
Tehran, Iran
ata.jodeiri@ut.ac.ir

Mohammad Mohsen Goodarzi
Department of Biomedical Engineering,
Buein Zahra Technical University,
Buein Zahra, Qazvin, Iran
mm.goodarzi@bzte.ac.ir

Shabnam Gholamdokht Firooz
School of Electrical & Computer Engineering
University of Tehran
Tehran, Iran
shabnam.firooz@ut.ac.ir

*Abstract*—Clipping, as a current nonlinear distortion, often occurs due to the limited dynamic range of audio recorders. It degrades the speech quality and intelligibility and adversely affects the performances of speech and speaker recognitions. In this paper, we focus on enhancement of clipped speech by using a fully convolutional neural network as U-Net. Motivated by the idea of image-to-image translation, we propose a declipping approach, namely "U-Net declipper" in which the magnitude spectrum images of clipped signals are translated to the corresponding images of clean ones. The experimental results show that the proposed approach outperforms other declipping methods in terms of both quality and intelligibility measures, especially in severe clipping cases. Moreover, the superior performance of the U-Net declipper over the well-known declipping methods is verified in additive Gaussian noise conditions.

## I. Introduction

Audio clipping is a common nonlinear distortion which degrades the hearing quality and Intelligibility. Clipping occurs in initial steps of audio recording, specifically in microphone and A/D blocks due to the insufficient dynamic range of them, and causes information loss.

In addition to degraded hearing quality and intelligibility, it has been shown that clipping adversely affects Automatic Speech Recognition [1], [2] and Speaker Recognition performances [3], [4]. So it is interested to revert the clipping distortion using a process called declipping. This process is usually performed by estimating clipped samples using reliable and unclipped samples around them.

Let $x(t)$ be a clean speech, the clipped speech $y(t)$ can be simulated by applying the clipping threshold $\theta$ to $x(t)$ as:

$$y(t) = \begin{cases} x(t), & \text{if } |x(t)| \leq \theta \\ \theta \cdot sgn(x(t)), & \text{if } |x(t)| > \theta \end{cases} \quad (1)$$

where $sgn(.)$ is the sign function. Fig. 1 shows a speech signal clipped by $\theta = 0.25$. With the above description, declipping goals to retrieve $x(t)$ from $y(t)$.

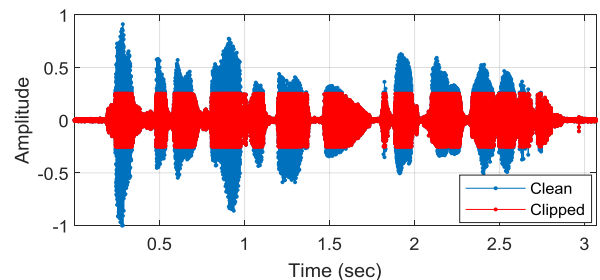

Fig. 1. Representation of speech clipped by $\theta = 0.25$.

Because clipping causes information loss, almost all techniques used for declipping involve constructing a model of clean and non-clipped signals to recover lost data. Their difference is in the model used and how the model is applied. Early studies employed simple parametric models such as AutoRegressive (AR) to interpolate clipped samples [5]. Statistical and generative models such as Monte Carlo filtering [6], [7] are also proposed to estimate distorted samples. But no reasonable quality was achieved until the introduction of sparsity-based declipping techniques [8], [9]. Sparse representation of speech signal has been successfully used in order to estimate noisy degraded feature of speech in automatic speech recognition tasks [10], [11]. These techniques utilize an inverse problem to represent frames of the audio signal with sparse weight vectors of predefined dictionary atoms. The clipped samples are marked as missing and estimated by applying sparse weight over the corresponding index of dictionary atoms. In [12], a regularized version of Iterative Hard Thresholding (IHT) [13] is proposed to improve declipping performance while reducing computational cost. The same algorithm is further developed to both sparse synthesis and sparse analysis scenarios [14]. The sparsity-based techniques reach their summit by using dictionary learning method proposed in [15]. A dictionary learned from clipped data and matched to reconstruction process, produced state-of-the-art results.

The process of reconstructing clipped samples could also be modeled using a mapping from clipped samples to original clean ones. The mapping used should be able to cover all speech variabilities. Such a powerful and comprehensive mapping brings Deep Neural Network (DNN) on the table. DNNs as a mapping of degraded/distorted speech to clean speech have been successfully proposed in many studies [16], [17]. Related to the speech distortion as clipping, the authors in [3] employed a DNN as a fully-connected (FC) network with 3 hidden layers for mapping a 9-frame window of MFCCs of clipped speech to a single frame of the clean MFCCs. However, they only utilized the MCEP features enhanced by DNN for a speaker recognition task and did not conduct any experimental analysis on the reconstruction of clipped speech.

In this study, by specifically focusing on the speech declipping task, we exploit a fully convolutional network (FCN) instead of FC networks which do not fully utilize the spectral structures of the speech. Moreover, instead of MCEP features, we apply spectrogram images as the input/output features to the FCN. Motivated by the recent success of a specific type of FCNs, namely U-Net [18], in many image-to-image translation tasks, we investigate it's applicability in a speech processing domain as speech declipping. In a simple word, declipping by U-Net can be interpreted as an image-to-image translation where the spectrogram image of the clipped speech is transformed to the spectrogram image of the clean one. Fig. 2 depicts the U-Net structure proposed for speech declipping.

The rest of the paper is organized as follows: Section II firstly describes the architecture of U-Net translator and then proposes the speech declipper based on U-Net. Section III details the experimental results of the proposed approach in comparison with other well-known speech declipping methods. Finally, conclusions are presented in Section IV.

## II. METHOD

### A. U-Net Architecture

U-net, as a fully convolutional neural network, is commonly used for image-to-image translation such as image segmentation and consists of three paths including contraction, expansion and skip connection. Novelty of the U-net is in concatenating the same-level features together and building richer feature maps by combining the local and global features extracted in contraction and expansion paths respectively.

In contraction path local and structural features are extracted. The input size of the image is reduced through the contraction path in order to increase the receptive field, make the model robust to noise and artifact and also decrease the computational cost. Increasing the receptive field leads to propagate global information in both time and frequency domains. The receptive field of the neurons is maximized in the bottleneck layer that causes each neuron to be influenced by the large area of the spectrum and for this reason, bottleneck proposes the richest feature maps.

Along with all the mentioned benefits of contraction path, losing the spatial resolution is the main drawback of that. In order to compensate for the resolution reduction problem, the contraction path is connected to the expansion path, making the U shape of the U-Net.

Connection between same-level features recovers the localization information. Skip connections combine the shallow, local and low-level features from encoder to the deep, semantic and high-level features from decoder. Skip-connection was demonstrated to recover full spatial resolution of the image [18]. In addition, the skip connections not only keep the network from gradient vanishing problem but also make easier to propagate the input information through the network in the training phase.

Contraction path follows the typical fully convolutional networks used for feature extraction in various machine vision tasks. It includes a sequence of two 3x3 CNNs with ReLU non-linearity functions followed by a 2x2 max-pooling layer with stride of 2. This sequence is repeated four times and after each time, the number of filters is doubled. Two CNNs in the bottleneck of the U-Net connect the contraction path to expansion path. The expansion path includes a sequence of two 3x3 CNNs with ReLU non-linearity functions followed by one up-sampling layer and one 2x2 CNN with ReLU non-linearity. This sequence is repeated four times and after each time, the number of filters is halved. Finally, a 1×1 CNN operation with sigmoid (or softmax for multi-label segmentation) function is performed to generate the final segmentation mask. In all four levels, the output of the CNN layer in contraction path is transferred to the corresponding level in the expansion path.

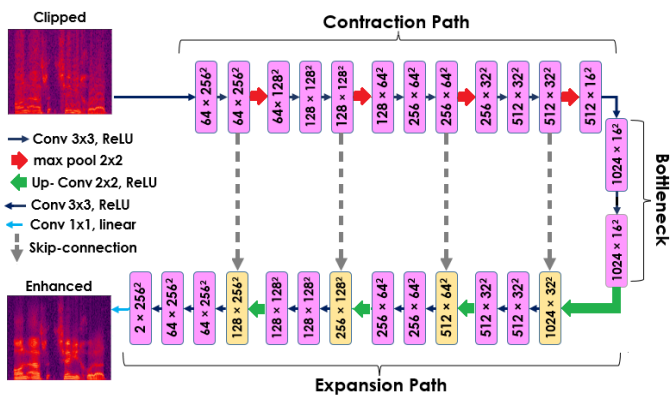

Fig. 2. U-Net structure proposed for speech declipping.

### B. U-Net Declipper

The proposed approach, namely U-Net declipper, consists of three stages as follows:

*Stage 1) Extracting spectrum images*

Firstly, we apply short-time Fourier transform (STFT) to the (clipped/clean) speech signal and compute the short-time feature vectors representing the logarithm of the magnitude spectrum (in short, log-mag spectrum). To compute STFT, the frame length, frame shift and the number of Fourier transform are selected as 32ms, 8ms and 512, respectively. As a result, the dimension of each log-mag spectrum vector is 257. In the following, spectrum images are made by concatenating 256 successive log-mag spectrum vectors. It's notable that in order to keep the image symmetry, the highest frequency bin is ignored. Consequently, each spectrum image is a 256x256 image representing the log-mag values at 256x256 time-frequency units. Note that before making spectrum images, only for clipped speech, mean-variance normalization is applied to the log-mag spectrum vectors at the speech utterance level. At the end of this stage, an image set, consisting of spectrum images of clipped signals and the corresponding clean ones, is provided.

*Stage 2) Training U-Net declipper*

The training spectrum image set provided from the previous stage is used to learn the U-Net model. The proposed model is implemented using Tensorflow library and optimized by minimizing Adam algorithm with 0.0002 learning rate for 50 epochs. At each epoch, a mini-batch size of 5 images is randomly selected from the training images and fed into the network to learn the parameters. Although in the image segmentation tasks, the softmax or sigmoid function is used at the end of the network for confining the output to the binary image, we use the linear function for generating enhanced spectrum image. Moreover, the mean square error (MSE) as the cost function is minimized during training.

*Stage 3) Reconstructing clipped speech*

At the signal reconstruction stage, firstly, the spectrum images of the input clipped speech are extracted as expressed at Stage 1. The input images are fed to the trained U-Net declipper to generate the enhanced spectrum images. Finally, the phase information of the clipped signal is combined with the enhanced log-mag spectrum and Inverse STFT is then used to generate the time-domain declipped speech.

## III. EXPERIMENTAL RESULTS

### A. Dataset Preparation

FARSDAT corpus [19] was utilized to provide the clipped dataset. FARSDAT is a Persian read speech composed of 6080 phonetically balanced sentences, 20 sentences uttered by each of 304 speakers. All sentences were randomly partitioned into three sets, namely training, development, and test, that respectively consist of about 75%, 10% and 15% of the whole corresponding corpus. There is no overlap between the speakers of these three sets. For generating the clipped speech, the clipping threshold ($\theta$) in (1), was selected to yield the corresponding signal to distortion ratio (SDR) as $SDR(x, y) = 10\log_{10}\left(\|x\|^2 / \|x-y\|^2\right)$. In order to prepare the clipped dataset, for each clean speech sentence in either training or development set, six clipped speech sentences were artificially generated by six SDRs from the set {1, 2, 5, 10, 15, 20} dB. To evaluate the SDR mismatch, each clean sentence in the test set was clipped by six SDRs from a different set as {0.5, 1.5, 3.5, 7.5, 12.5, 17.5} dB.

### B. Performance Measures

We evaluate the performance of the declipping methods based on three objective measures representing both quality and intelligibility of enhanced speech. Two criteria of Perceptual Evaluation of Speech Quality (PESQ) [20] and Log Likelihood Ratio (LLR) [21] are considered as speech quality measures that adopt the values in the intervals of [-0.5,4.5] and [0,2], respectively. The Extended Short-Time Objective Intelligibility (ESTOI) [22] is related to the intelligibility and limited to 0 to 1. For two measures of PESQ and ESTOI, higher values are better, but for LLR, the lower is better.

### C. Results and Discussion

We considered four well-known declipping methods as IHT [13], Dictionary Learning (DL) [15], Consistent-IHT [12], and Consistent-DL [15] to compare with the proposed one. Tables I to III demonstrate the comparisons of these methods on the basis of PESQ, ESTOI and LLR measures, respectively. The results on the clipped case were also shown. In these tables, we represented the average performances for different clipping thresholds, ranging from heavy clipping (SDR=0.5 dB) to light one (SDR=17.5 dB). Finally, the total average on all six clipping SDRs for each method was also shown.

Table I Comparisons of methods based on PESQ measure

| Method | SDR (dB) | | | | | | |
|---|---|---|---|---|---|---|---|
| | 0.5 | 1.5 | 3.5 | 7.5 | 12.5 | 17.5 | Avg. |
| Clipped | 1.12 | 1.2 | 1.42 | 2.00 | 2.82 | 3.48 | 2.00 |
| IHT | 1.03 | 1.03 | 1.05 | 1.18 | 1.74 | 2.62 | 1.44 |
| DL | 1.03 | 1.03 | 1.05 | 1.17 | 1.72 | 2.63 | 1.44 |
| Cons-IHT | 1.13 | 1.26 | 1.6 | 2.42 | 3.23 | 3.69 | 2.22 |
| Cons-DL | 1.16 | 1.3 | 1.66 | 2.51 | 3.39 | **3.89** | 2.32 |
| Proposed | **2.07** | **2.19** | **2.46** | **2.97** | **3.49** | 3.79 | **2.83** |

The best result in each condition was marked in bold.

Table II Comparisons of methods based on ESTOI measure

| Method | SDR (dB) | | | | | | |
|---|---|---|---|---|---|---|---|
| | 0.5 | 1.5 | 3.5 | 7.5 | 12.5 | 17.5 | Avg. |
| Clipped | 0.59 | 0.65 | 0.75 | 0.87 | 0.94 | 0.97 | 0.79 |
| IHT | 0.04 | 0.12 | 0.33 | 0.66 | 0.86 | 0.94 | 0.49 |
| DL | 0.04 | 0.12 | 0.33 | 0.65 | 0.86 | 0.94 | 0.49 |
| Cons-IHT | 0.59 | 0.68 | 0.78 | 0.90 | 0.96 | 0.98 | 0.82 |
| Cons-DL | 0.61 | 0.70 | 0.81 | **0.91** | **0.97** | **0.99** | 0.83 |
| Proposed | **0.78** | **0.81** | **0.85** | 0.90 | 0.93 | 0.94 | **0.87** |

The best result in each condition was marked in bold.

Table III Comparisons of methods based on LLR measure

| Method | SDR (dB) | | | | | | |
|---|---|---|---|---|---|---|---|
| | 0.5 | 1.5 | 3.5 | 7.5 | 12.5 | 17.5 | Avg. |
| Clipped | 0.81 | 0.63 | 0.44 | 0.24 | 0.11 | 0.05 | 0.38 |
| IHT | 1.38 | 1.15 | 0.87 | 0.49 | 0.22 | 0.09 | 0.70 |
| DL | 1.39 | 1.18 | 0.91 | 0.51 | 0.21 | 0.09 | 0.72 |
| Cons-IHT | 0.65 | 0.50 | 0.35 | 0.19 | 0.10 | 0.05 | 0.31 |
| Cons-DL | 0.74 | 0.56 | 0.36 | 0.18 | **0.08** | **0.03** | 0.33 |
| Proposed | **0.40** | **0.32** | **0.25** | **0.16** | 0.11 | 0.08 | **0.22** |

The best result in each condition was marked in bold.

Given Tables I-III, the following results can be expressed:

- All measures are sensitive to the existence of clipping. At a glimpse, the quality of clipped speech (the first rows of the tables) decreases when clipping threshold is increasing.
- Given the PESQ quality measure, the proposed method has resulted in the highest values for all test SDRs except 17.5 dB SDR in which Consistent-DL is the best.
- According to the ESTOI intelligibility measure, for severe clippings such as 0.5 and 1.5 dB, the proposed approach considerably outperforms the other methods. However, in light clipping cases such as 12.5 and 17.5 dB, both Consistent-DL and Consistent-IHT are better than the suggested one.
- Regarding LLR quality measure, the proposed method has desirably yielded the lower values for hard and moderate SDRs. Although, for light clips, Consistent-DL is still better.
- Give the average performances on all six clipping levels, the proposed method shows the best performances in terms of all three measures.

Fig. 3 shows the spectrograms representing the reconstructions of a clipped speech (SDR=1.5 dB) with the Consistent-DL and the proposed methods. We can see that the clipping leads to attenuate the energy of main harmonics and generally distort the harmonic structure. Compared with the reconstruction achieved by Consistent-DL, the proposed approach has desirably amplified the energy at main harmonics and produced the much more smooth harmonic structure.

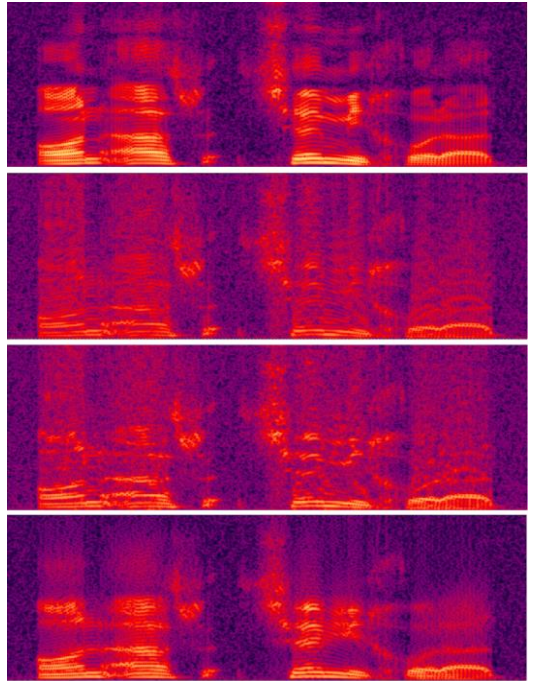

Fig. 3. Spectrogram representations of clean speech, clipped one, and those achieved by Consistent-DL and proposed methods, respectively (from top to down).

Fig. 4 illustrates the declipping performances for speech signals smeared with additive Gaussian noise with variance $\sigma^2$ and clipped at SDR=3.5 dB. According to Tables I-III, we selected the Consistent-DL and Consistent-IHT, as the two much better declipping methods, for comparing with the presented one. It can be observed that the proposed declipping has considerably improved PESQ and LLR measures for all six noise levels, while the two compared methods have yielded negligible enhancements. According to the ESTOI values, the proposed method is still the best for noise levels below 0.05, but Consistent-DL has resulted in the highest ESTOI for severe noises. Finally, note that all experiments were conducted on a workstation with Core i7 and 3.4 GHz CPU with 8GB RAM and Nvidia GeForce GTX 1060 GPU.

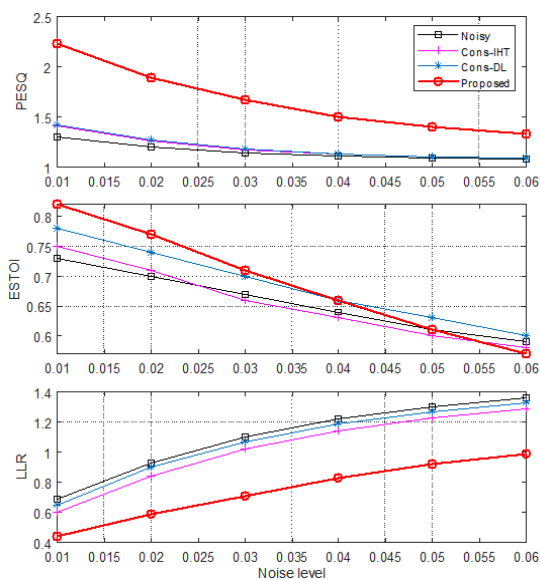

Fig. 4. Performance comparisons in additive Gaussian noise conditions.

## IV. CONCLUSION

The present study aimed to introduce a novel speech declipper which generates the enhanced speech signals with high speech qualities and intelligibilities. To do so, we took advantages of image-to-image translation networks presented in image processing domains such as U-Net. The prominent characteristic of U-Net is in building richer feature maps by combining the shallow, local and low-level features from the encoder part with the deep, semantic and high-level features from the decoder. Given this fact, we proposed U-Net declipper which translates the log-magnitude spectrum images extracted from clipped speech signals to the corresponding images from clean ones. Visual results exhibited that this translation desirably preserves the smoothness of harmonic structures in the enhanced spectrum image. Moreover, from both perspectives of speech quality and intelligibility measures, such as PESQ, LLR and ESTOI, the proposed declipper outperforms the other well-known declipping methods. Surprisingly, the superior outcomes by the proposed approach were more pronounced as the clipping level increased. As a future work, we plan to apply the output of the proposed declipper to a speech recognition application.